\documentclass[aps,prl,superscriptaddress,amsfonts,amsmath,amssymb,showpacs,floatfix,reprint]{revtex4-1}

\usepackage{url}
\usepackage{bm}
\usepackage{graphicx}
\usepackage{amsmath}
\usepackage{amstext}
\usepackage{amssymb}
\usepackage{amsfonts}
\usepackage{amsbsy}
\usepackage{verbatim}
\usepackage{color}
\usepackage[colorlinks=true, urlcolor=blue, linkcolor=blue, citecolor=blue, pdftex]{hyperref}
\usepackage{multirow}
\usepackage{pdfpages}
\usepackage{floatrow}
\usepackage{gensymb}
\usepackage{textcomp}
\usepackage{enumitem}
\usepackage[version=3]{mhchem}

\def\q0{{(q,0)}}
\def\p0{{(\pi,0)}}

\begin{document}

\title{Intertwined nematic orders in a frustrated ferromagnet}

\author{Yasir Iqbal}
\email[]{yiqbal@physik.uni-wuerzburg.de}
\affiliation{Institute for Theoretical Physics and Astrophysics, Julius-Maximilian's University of W{\"u}rzburg, Am Hubland, D-97074 W{\"u}rzburg, Germany}
\author{Pratyay Ghosh}
\affiliation{School of Physical Sciences, Jawaharlal Nehru University, New Delhi 110067, India}
\author{Rajesh Narayanan}
\affiliation{Department of Physics, Indian Institute of Technology Madras, Chennai 600036, India}
\affiliation{Asia Pacific Center for Theoretical Physics (APCTP), Pohang, Gyeongbuk, 790-784, South Korea}
\author{Brijesh Kumar}
\affiliation{School of Physical Sciences, Jawaharlal Nehru University, New Delhi 110067, India}
\author{Johannes Reuther}
\affiliation{Dahlem Center for Complex Quantum Systems and Fachbereich Physik, Freie Universit{\"a}t Berlin, D-14195 Berlin, Germany}
\affiliation{Helmholtz-Zentrum Berlin f{\"u}r Materialien und Energie, D-14109 Berlin, Germany}
\author{Ronny Thomale}
\affiliation{Institute for Theoretical Physics and Astrophysics, Julius-Maximilian's University of W{\"u}rzburg, Am Hubland, D-97074 W{\"u}rzburg, Germany}

\date{\today}

\begin{abstract}
We investigate the quantum phases of the frustrated spin-$\frac{1}{2}$ $J_1$-$J_2$-$J_3$ Heisenberg model on the square lattice with ferromagnetic $J_1$ and antiferromagnetic $J_2$ and $J_3$ interactions. Using the pseudo-fermion functional renormalization group technique, we find an intermediate paramagnetic phase located between classically ordered ferromagnetic, stripy antiferromagnetic, and incommensurate spiral phases. We observe that quantum fluctuations lead to significant shifts of the spiral pitch angles compared to the classical limit. By computing the response of the system with respect to various spin rotation and lattice symmetry-breaking perturbations, we identify a complex interplay between different nematic spin states in the paramagnetic phase. While retaining time-reversal invariance, these phases either break spin-rotation symmetry, lattice-rotation symmetry, or a combination of both. We therefore propose the $J_1$-$J_2$-$J_3$ Heisenberg model on the square lattice as a paradigmatic example where different intimately connected types of nematic orders emerge in the same model.
\end{abstract}

\maketitle

%%%%%%%%%%%%%%%%%%%%%%%%%%%%%%%%%%%%%%%%%%%%
\section{Introduction}
A cardinal theme in modern condensed matter physics is the search for novel states of matter, such as quantum spin liquids. Their identification as a host of fractional spin excitations and topological order~\cite{Balents-2010}, and the seminal works of Anderson~\cite{Anderson-1987,Baskaran-1987} highlighting the possible connection of the ``resonating valence-bond'' scenario to high-$T_{c}$ have established the study of spin liquids as one of the most active areas of research. The traditional recipe for obtaining spin liquids involves the task of melting magnetic long-range order by either geometrical or parametrical frustration in a quantum antiferromagnet. Recently, the synthesis of a growing number of quantum magnets with competing antiferromagnetic (AF) and ferromagnetic (FM) interactions, has also fuelled the search for paramagnets in a FM environment~\cite{balz2016,Bernu2013,Fak2012,Bieri-2015,Iqbal-2015,Balents-2016}.
While frustration from the interplay between AF and FM couplings can be similarly efficient in melting magnetic order as in the purely AF case, the propensity for resonating singlet bonds is weakened in favor of resonating {\it triplet} bonds. The latter scenario opens up the possibility of stabilizing an exotic variant of a quantum paramagnet, called a {\it spin nematic}~\cite{Andreev-1984,Gorkov-1990,Podolsky-2005}. While a spin nematic is characterized by an absence of dipolar magnetic order, i.e., $\langle \hat{{\bf S}}_{i} \rangle=0$ (where $\hat{\mathbf{S}}_{i}$ denotes the spin operator at site $i$), and respects time-reversal symmetry, it breaks SU$(2)$ spin-rotation symmetry due to a nonzero quadrupolar order parameter of the form $\mathcal{O}_{ij}^{\mu\nu}=\langle\hat{S}_i^\mu \hat{S}_j^\nu\rangle-\frac{\delta_{\mu\nu}}{3}\langle\hat{\bf S}_i\cdot\hat{\bf S}_j\rangle$ (with $\mu$, $\nu=x,y,z$). It can be viewed as a quantum spin analog of the nematic state in liquid crystals where the spin direction takes the role of the orientation of molecules. Historically, the search has focused on $S=1$ Heisenberg systems with additional biquadratic [$(\mathbf{\hat{S}}_{i}\cdot \mathbf{\hat{S}}_{j})^{2}$] interactions~\cite{Blume-1969,Matveev-1973,Matveev-1983,Niko-1988,Chubukov-1990,Chandra-1990b,Chubukov-1991,Chandra-1991,Penc-2010,Bieri-2012}, and more recently on systems in a magnetic field~\cite{Picot-2015,Seabra-2016}. On the experimental front, the detection of quadrupolar order is extremely challenging, e.g., in thermodynamic measurements their response is widely indistinguishable from that of antiferromagnets~\cite{Andreev-1984}. However, techniques such as neutron scattering in a magnetic field have been suggested as probes to detect nematic order~\cite{Gorkov-1993}. Only very recently has the possible existence of nematic orders been reported in iron-based high-$T_{c}$ superconductors such as FeSe~\cite{Glasbrenner-2015,Wang-2015,Gong-2016,Hu-2016}, in the mineral linarite~\cite{Willenberg-2016}, and in ultra-cold atomic gases~\cite{Hamley-2012} (see Ref.~\cite{frust} for further details on candidate materials).

%==========QPD==================
\begin{figure*}
\includegraphics[width=1.0\textwidth]{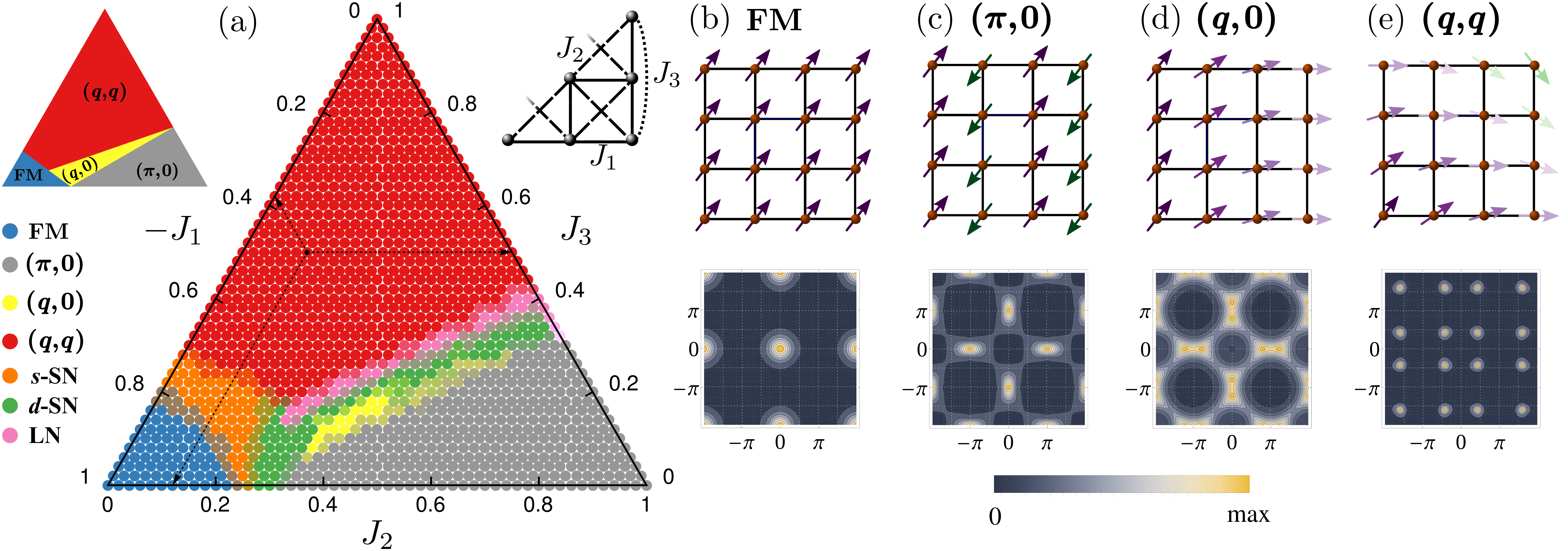}
\caption{\label{fig:fig1}
(a) Quantum phase diagram of the spin-$\frac{1}{2}$ $J_1$-$J_2$-$J_3$ Heisenberg model [Eq.~(\ref{Hamil})] obtained within PFFRG. The coupling constants are normalized such that $|J_1|+J_2+J_3=1$. The phase diagram hosts a large PM domain featuring $s$-SN, $d$-SN, and LN orders. Interpolating colors indicate regions of uncertainties near the phase boundaries. The corresponding classical phase diagram is shown in the upper left. A depiction of the exchange couplings is shown in the upper right. (b)${-}$(e) Illustrations of the real space pattern (upper row) and momentum-space resolved magnetic susceptibility profile (lower row) in units of $1/|J_{1}|$ for magnetism at ordering vectors ${\bf Q}=(0,0),~(\pi,0),~(q,0)$, and $(q,q)$, evaluated at the parameters $(|J_{1}|,J_{2},J_{3})=(0.84,0.08,0.08),~(0.52,0.42,0.06),~(0.48,0.38,0.14),~(0.52,0.06,0.42)$, respectively.}
\end{figure*}
%================================

In this paper, we investigate a simple frustrated 2D system with competing AF and FM interactions being the spin-$\frac{1}{2}$ Heisenberg model on the square lattice with FM nearest-neighbor ($J_{1}$), and AF second neighbor ($J_{2}$) couplings. Strong frustration in the vicinity of the classical ($S\to\infty$) transition point at $J_2/|J_1|=1/2$ separating ferromagnetic order at $J_2/|J_1|<1/2$ [with wave vector ${\bf Q}_\text{cl}=(0,0)$, see Fig.~\ref{fig:fig1}(b)] from collinear stripe AF-order at $J_2/|J_1|>1/2$ [with ${\bf Q}_\text{cl}=(\pi,0)$ or $(0,\pi)$, see Fig.~\ref{fig:fig1}(c)] has raised the question of an intermediate paramagnetic phase in the quantum case, which could possibly be nematic in nature. Herein, we study different nematic phases on the square lattice in a broader context by adding an AF third-neighbor interaction $J_3$. Classically, this gives rise to the appearance of two additional magnetic states~\cite{Rastelli-1979}, a 1D helimagnet (HM) consisting of lines of parallel spins in the $(0,1)$ or $(1,0)$ direction [with ${\bf Q}_\text{cl}=(\pm q,0)$ or $(0,\pm q)$, see Fig.~\ref{fig:fig1}(d)] and a 2D helimagnet consisting of lines of parallel spins in the $(1,1)$ or $(1,-1)$ direction, [with ${\bf Q}_\text{cl}=(\pm q,\pm q)$, see Fig.~\ref{fig:fig1}(e) and the inset of Fig.~\ref{fig:fig1}(a) for the classical phase diagram]. In general, these helical orders are incommensurate with the lattice periodicity. Little is known about the effects of quantum fluctuations in this model. In the lowest (first) order in $1/S$, a significant enhancement of the stripe AF phase at the expense of FM and HM states has been reported~\cite{Rastelli-1986,Rastelli-1983,Rastelli-1985,Dmitriev-1997}, and the effect of the $J_{3}$ interaction was also analyzed using exact-diagonalization (ED) on systems up to $36$ spins~\cite{Sindzingre-2009,Sindzingre-2010}. 

To shed more light on the quantum effects in the $J_{1}$-$J_{2}$-$J_{3}$ Heisenberg model on the square lattice, we employ a state-of-the-art implementation of the pseudo-fermion functional renormalization group (PFFRG) method enabling access to large correlation areas ($\sim1000$-sites). In particular, we introduce generalized nematic response functions within the PFFRG framework. Aside from $d$-wave spin nematic ($d$-SN) order that breaks SU$(2)$ spin-rotation symmetry as well as lattice-rotation symmetry~\cite{Shannon-2006,Sindzingre-2007,Sindzingre-2009,Sindzingre-2010}, we also find regimes of $s$-wave spin-nematic ($s$-SN) order which exclusively break spin-rotation symmetry (while keeping lattice symmetries intact), and lattice nematic (LN) orders which only break lattice-rotation symmetries (while keeping spin-rotation symmetries intact). Our main results are summarized as follows: quantum fluctuations melt significant portions of the HM and FM orders, stabilizing a PM phase over a vast region of parameter space. The PM phase features different domains wherein either the $d$-SN, $s$-SN, or LN response function dominates, indicating that all three types of nematic orders might be realized in the system.

\section{Model and method}
The Hamiltonian of the $J_1$-$J_2$-$J_3$ Heisenberg model reads

\begin{equation}\label{Hamil}
\hat{{\cal H}} = J_1 \sum_{{\langle i,j \rangle}} \mathbf{\hat{S}}_{i} \cdot \mathbf{\hat{S}}_{j}
+J_2 \sum_{{\langle\langle i,j \rangle\rangle}} \mathbf{\hat{S}}_{i} \cdot \mathbf{\hat{S}}_{j}
+J_3 \sum_{{\langle\langle\langle i,j \rangle\rangle\rangle}} \mathbf{\hat{S}}_{i} \cdot \mathbf{\hat{S}}_{j}\,,
\end{equation}
where $J_{1}\leqslant0$ (FM) and $J_{2},J_{3}\geqslant0$ (AF) and ${\langle i,j \rangle}$, ${\langle\langle i,j \rangle\rangle}$, and ${\langle\langle\langle i,j \rangle\rangle\rangle}$ denote sums over nearest-neighbor (NN), second-nearest-neighbor (2-NN), and third-nearest-neighbor (3-NN) pairs of sites, respectively [see inset of Fig.~\ref{fig:fig1}(a)]. Within the PFFRG scheme~\cite{Reuther2010,Thomale-2011,Abanin-2011,Reuther2011,Reuther2011_2,Reuther2014} this Hamiltonian is first rewritten in terms of Abrikosov pseudofermions, $\mathbf{\hat{S}}_{i}=\frac{1}{2}\sum_{\alpha,\beta}\hat{c}^{\dagger}_{i,\alpha}\pmb{\sigma}_{\alpha\beta}\hat{c}_{i,\beta}$, where $\alpha$, $\beta=\uparrow$ or $\downarrow$, and $\hat{c}^{\dagger}_{i,\alpha}$ ($\hat{c}_{i,\alpha}$) are the pseudofermion creation (annhilation) operators, and $\pmb{\sigma}$ is the Pauli vector. The introduction of pseudofermions is associated with an enlargement of the Hilbert space, which, in addition to the physical spin-$\frac{1}{2}$ states, also contains spurious empty or doubly occupied states carrying zero spin. Since such unphysical occupations effectively act like a vacancy in the spin lattice, they are associated with a finite excitation energy of the order of the exchange couplings. This can be tested by adding onsite terms $\sim\sum_i {\bf S}_i^2$ to the Hamiltonian which change the energetic difference between physical and unphysical occupations\footnote{Maria L. Baez, J. Reuther, in preparation}. As a consequence, the ground state of the fermionic system probed within PFFRG is identical to the ground state of the original spin model where each site is singly occupied.

Following the introduction of an infrared cutoff $\Lambda$ along the Matsubara frequency axis in the fermion propagator, the FRG ansatz is formulated in terms of an exact but infinite hierarchy of coupled flow equations for the $m$-particle vertex functions~\cite{Metzner2012,Platt2013}. For a numerical implementation, the hierarchy of equations is truncated to keep only the self-energy and two-particle vertex functions. This truncation is performed such that via self-constistent feedback of the self-energy into the two-particle vertex, the approach remains separately exact in the large $S$ limit as well as in the large $N$ limit [where the spins' symmetry group is promoted from SU$(2)$ to SU$(N)$]. This property allows for an unbiased investigation of the competition between magnetic ordering tendencies and quantum paramagnetic behavior. Approximations in the PFFRG scheme concern subleading orders in $1/S$ and $1/N$ such as three-particle vertices. Deep inside magnetically ordered phases the exactness of the PFFRG in the leading order in $1/S$ ensures that classical magnetic states are correctly captured. On the other hand, the leading $1/N$ terms guarantee a proper description of nonmagnetic states deep inside magnetically disordered phases. The neglected subleading terms given by fermionic three-particle vertices can become important near quantum critical points which are characterized by a competition between quantum fluctuations and ordering tendencies. As a consequence, phase transitions are always subject to an uncertainty within PFFRG. Three-particle vertices can also become important in chiral spin liquids where they describe chiral order parameters of the form $\sim\langle ({\bf S}_i\times{\bf S}_j)\cdot{\bf S}_k\rangle$. Therefore, the current implementation of the PFFRG does not resolve the propensity of a spin system to form chiral spin liquids.

The two-particle vertex in real space is related to the static (imaginary time-integrated) spin correlator
\begin{equation}
C_{ij}^{\mu\nu}=\int_0^\infty d\tau\langle\hat{S}_i^\mu(\tau)\hat{S}_j^\nu(0)\rangle\label{correlator}
\end{equation}
with $\hat{S}_i^\mu(\tau)=e^{\tau\hat{\mathcal{H}}}\hat{S}_i^\mu e^{-\tau\hat{\mathcal{H}}}$. Within PFFRG, the thermodynamic limit is approximated by calculating the correlators $C^{\mu\nu}_{ij}$ only up to a maximal distance between sites $i$ and $j$. The main physical outcome of the PFFRG are the Fourier-transformed correlators, i.e., the static susceptibility $\chi^{\mu\nu,\Lambda}({\bf q})$ evaluated as a function of the RG scale $\Lambda$. After performing the Fourier-transform, we generally have access to a continuous range of {\bf q} vectors within the Brillouin zone. However, since correlations beyond a certain maximal length are neglected, the Fourier sums contain a finite number of harmonics and sudden changes of the susceptibility in {\bf q} space can only be resolved with a limited accuracy. In our case, 15 lattice spacings corresponding to a total area of $31^{2}=961$ correlated sites yield well converged results and ensure a proper {\bf q}-space resolution. If a system develops magnetic order, the corresponding two-particle vertex channel anomalously grows upon decreasing $\Lambda$ and eventually causes the flow to become unstable as the channel flows towards strong coupling. Otherwise, a smooth flow behavior of the susceptibility indicates the absence of magnetic order. For further details about the PFFRG procedure, we refer the reader to Refs.~\cite{Reuther2010,Reuther2011,Reuther2011_2,Iqbal-2016}.

%=================PM===================
\begin{figure}
\includegraphics[width=1.0\columnwidth]{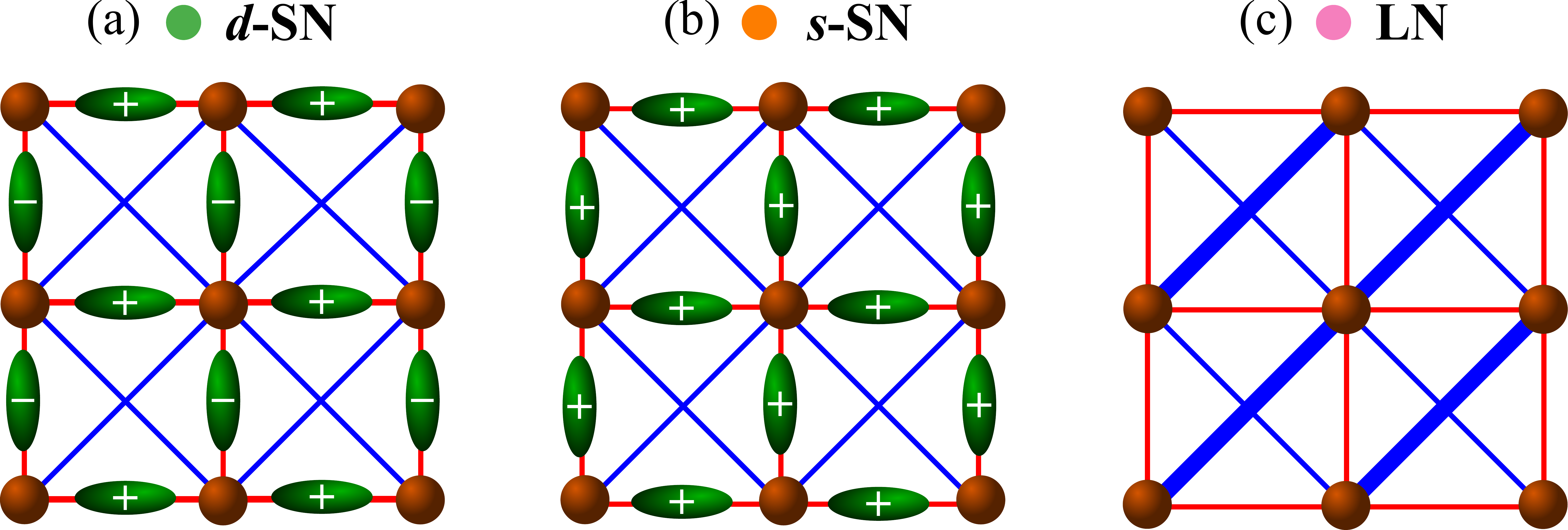}
\caption{\label{bias}
Illustration of the biasing patterns. (a) and (b) The oval with $+$($-$) sign represents a perturbation proportional to $\left|\sum_i(\hat{S}_i^z\hat{S}_{i+\hat{e}}^z-\hat{S}_i^x\hat{S}_{i+\hat{e}}^x-\hat{S}_i^y\hat{S}_{i+\hat{e}}^y)\right|$ with a positive (negative) prefactor and the respective nearest neighbor lattice vector $\hat{e}=\hat{x},\hat{y}$. (c) The perturbation is proportional to the difference in the thickness of the two diagonals, i.e., $\langle\hat{\bf S}_i\cdot\hat{\bf S}_{i+\hat{x}+\hat{y}}\rangle-\langle\hat{\bf S}_i\cdot\hat{\bf S}_{i+\hat{x}-\hat{y}}\rangle$.}
\end{figure}
%====================================

To probe the nature of the quantum paramagnetic phase, we examine nematic response functions of three different types of nematic states, the $d$-SN, $s$-SN, and LN. The corresponding order parameters $\mathcal{O}_{d\text{-SN}}$, $\mathcal{O}_{s\text{-SN}}$, and $\mathcal{O}_\text{LN}$ are given by
\begin{align}
\mathcal{O}_{d\text{-SN}}&=\mathcal{O}^{zz}_{i,i+\hat{x}}-\mathcal{O}^{xx}_{i,i+\hat{x}}=-(\mathcal{O}^{zz}_{i,i+\hat{y}}-\mathcal{O}^{xx}_{i,i+\hat{y}})\,,\notag\\
\mathcal{O}_{s\text{-SN}}&=\mathcal{O}^{zz}_{i,i+\hat{x}}-\mathcal{O}^{xx}_{i,i+\hat{x}}=\mathcal{O}^{zz}_{i,i+\hat{y}}-\mathcal{O}^{xx}_{i,i+\hat{y}}\,,\notag\\
\mathcal{O}_{\text{LN}}&=\langle\hat{\bf S}_i\cdot\hat{\bf S}_{i+\hat{x}+\hat{y}}\rangle-\langle\hat{\bf S}_i\cdot\hat{\bf S}_{i+\hat{x}-\hat{y}}\rangle\,,\label{orders}
\end{align}
where $\hat{x}$ and $\hat{y}$ denote unit vectors along the lattice directions and with the triplet order parameter $\mathcal{O}$ as given in the introduction. Furthermore, we assume that spin isotropy is always retained for spin rotations in the $x$-$y$ plane such that $\mathcal{O}^{xx}_{ij}=\mathcal{O}^{yy}_{ij}$. Due to the difference between spin correlations with $x$ and $z$ components, the $d$-SN and $s$-SN both break SU$(2)$ spin-rotation symmetry down to U$(1)$. Additionally, $d$-SN breaks the lattice-point group $C_{4v}$ down to $C_{2v}$, which is indicated by a relative minus sign between correlations along the $\hat{x}$ and $\hat{y}$ directions in the first line of Eq.~(\ref{orders}), leading to an effective $d$-wave character of this state. It is worth noting that the $d$-SN and $s$-SN order parameters are both of symmetric $n$-type, obeying $\mathcal{O}_{ij}^{\mu\nu}=\mathcal{O}_{ji}^{\mu\nu}$. This contrasts with the antisymmetric, chiral $p$-type nematic state where the order parameter is of the form $\mathcal{O}^\mu_{p,ij}=\epsilon_{\mu\nu\sigma}\mathcal{O}_{ij}^{\nu\sigma}$ and is argued to be stabilized in the presence of additional ring-exchange terms~\cite{Lauchli-2005}. Finally, the LN order parameter breaks the same lattice symmetries as the $d$-SN state but keeps SU$(2)$ spin-rotation symmetry intact. The LN state can therefore not be described by the triplet order parameters $\mathcal{O}^{\mu\nu}_{ij}$ but is probed by singlet spin-expectation values $\langle\hat{\bf S}_i\cdot\hat{\bf S}_{j}\rangle$~\cite{Iqbal-2016a}. Note that in Eq.~(\ref{orders}), the LN order parameter has been written as a difference between correlations along the diagonal $\hat{x}+\hat{y}$ and $\hat{x}-\hat{y}$ directions. As described below, this type of order parameter turns out to be particularly suitable to probe the LN state as compared to the nearest-neighbor term $\mathcal{O}'_{\text{LN}}=\langle\hat{\bf S}_i\cdot\hat{\bf S}_{i+\hat{x}}\rangle-\langle\hat{\bf S}_i\cdot\hat{\bf S}_{i+\hat{y}}\rangle$. For an illustration of the order parameters, see Fig.~\ref{bias}.

In general, the formation of a spin-nematic state is accompanied by a divergence of the corresponding order-parameter {\it susceptibility}, which is given by a {\it four}-spin correlator. In pseudo-fermion language such a correlator is represented by the fermionic four-particle vertex. The computation of such vertices is, however, far beyond the scope of current FRG implementations. We therefore pursue a simpler and more direct approach to probe the system with respect to these types of order. Collecting the operator content of the order parameters we define the perturbations
\begin{align}
\hat{\mathcal{H}}_{d\text{-SN}}&=\delta\sum_i(\hat{S}_i^z\hat{S}_{i+\hat{x}}^z-\hat{S}_i^x\hat{S}_{i+\hat{x}}^x-\hat{S}_i^y\hat{S}_{i+\hat{x}}^y)-(\hat{x}\rightarrow\hat{y})\,,\notag\\
\hat{\mathcal{H}}_{s\text{-SN}}&=\delta\sum_i(\hat{S}_i^z\hat{S}_{i+\hat{x}}^z-\hat{S}_i^x\hat{S}_{i+\hat{x}}^x-\hat{S}_i^y\hat{S}_{i+\hat{x}}^y)+(\hat{x}\rightarrow\hat{y})\,,\notag\\
\hat{\mathcal{H}}_{\text{LN}}&=\delta\sum_i(\hat{\bf S}_i\cdot\hat{\bf S}_{i+\hat{x}+\hat{y}}-\hat{\bf S}_i\cdot\hat{\bf S}_{i+\hat{x}-\hat{y}})\,.\label{order_ham}
\end{align}
Setting $0<\delta\ll|J_1|$, $J_2$, $J_3$ and adding these terms to $\hat{\mathcal H}$ [Eq.~(\ref{Hamil})] induces a small bias towards the respective type of nematicity, see Fig.~\ref{bias}. In PFFRG, the response of the system to these perturbations can be probed via the spin-spin correlator $C_{ij}^{\mu\nu}$ defined in Eq.~(\ref{correlator}). For the three nematic states, the biasing patterns lead to strengthened (weakened) correlators $C_+$ ($C_-$) along the respective spin directions/bonds given by
\begin{align}
d\text{-SN:\;}&C_+=\frac{1}{2}(C_{i,i+\hat{x}}^{zz}+C_{i,i+\hat{y}}^{xx})\,,\;C_-=\frac{1}{2}(C_{i,i+\hat{y}}^{zz}+C_{i,i+\hat{x}}^{xx}),\notag\\
s\text{-SN:\;}&C_+=C_{i,j}^{zz}\;,\;C_-=C_{i,j}^{xx}\,,\notag\\
{\rm LN}:\;&C_+=C^{\mu\mu}_{i,i+\hat{x}+\hat{y}}\,,\;C_-=C^{\mu\mu}_{i,i+\hat{x}-\hat{y}}\,,\label{cpcm}
\end{align}
where $i,j$ denote arbitrary nearest neighbors and $\mu=x,y,z$ can be any spin direction. Since $\hat{\mathcal{H}}_{d\text{-SN}}$ generates two inequivalent types of strengthened and weakened bonds we take the average in the first line of Eq.~(\ref{cpcm}). The generalized nematic responses $\kappa_\text{nem}$ are then defined by 
\begin{equation} 
\kappa_\text{nem}=\frac{J}{\delta}\frac{C_+^\Lambda - C_-^\Lambda}{C_+^\Lambda + C_-^\Lambda}\;,\label{nematic_susceptibility}
\end{equation}  
where $J$ is the coupling on the respective unperturbed bond. Note that Eq.~(\ref{nematic_susceptibility}) is normalized such that $\kappa_\text{nem}>1$ ($\kappa_\text{nem}<1$) corresponds to an enhancement (rejection) of the perturbation during the RG flow. 

%============Shift=================
\begin{figure}
\includegraphics[width=1.0\columnwidth]{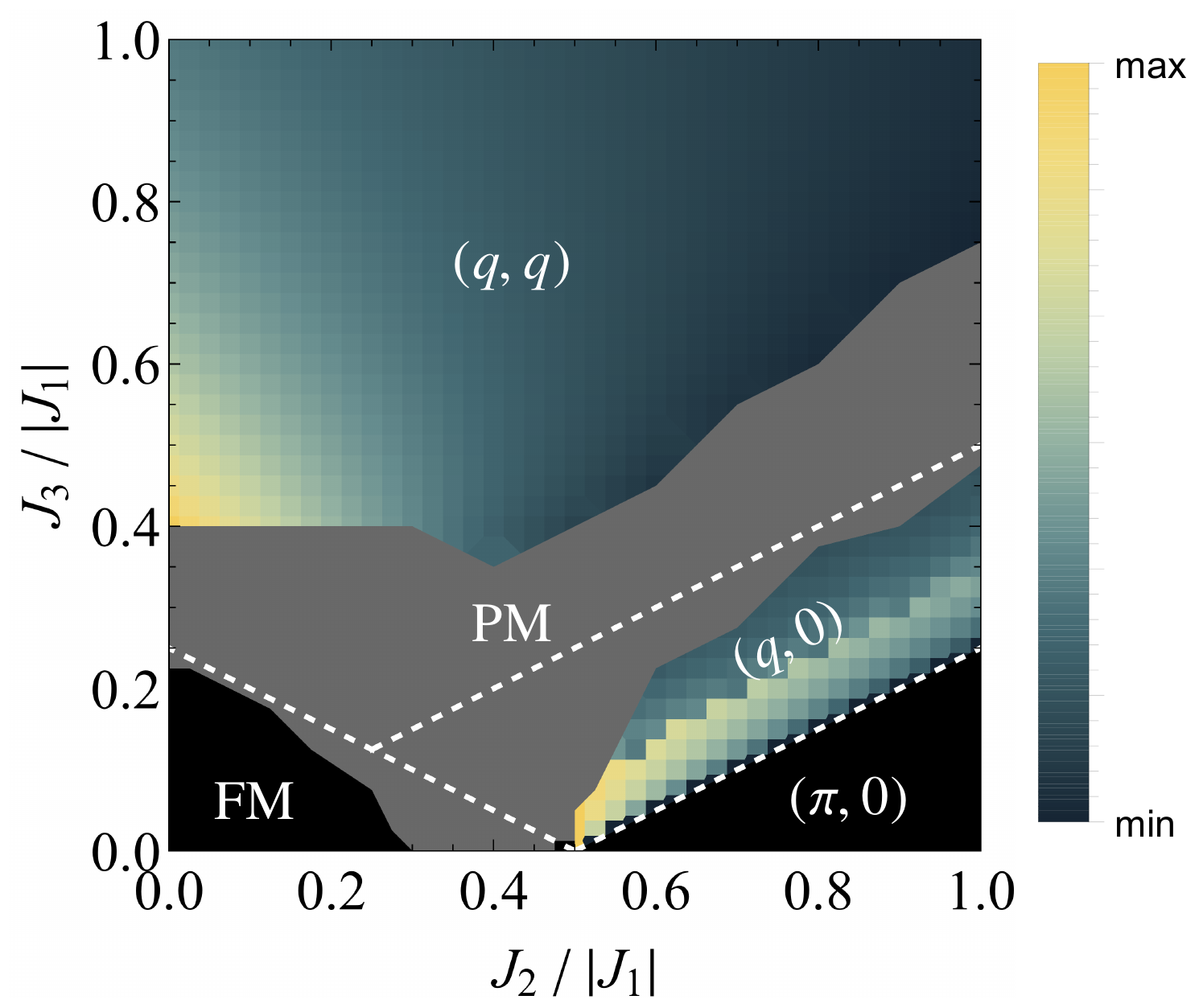}
\caption{\label{shift}
Deviation $\delta {\bf Q}={\bf Q}-{\bf Q}_\text{cl}$ of the ordering wave vector ${\bf Q}$ from its classical value ${\bf Q}_{\text{cl}}$ as a function of $J_{2}/|J_{1}|$ and $J_{3}/|J_{1}|$. The shifts in the black regions are identically zero, and the gray region denotes the PM phase. The classical boundaries are marked with white dashed lines. The maximum shifts in the $(q,q)$  and $(q,0)$ HM phases are $\approx 37\%$ and $\approx 100\%$, respectively, of the classical values.} 
\end{figure}
%================================

\section{Results}
The PFFRG quantum phase diagram of the spin-$\frac{1}{2}$ $J_{1}$-$J_{2}$-$J_{3}$ Heisenberg model of Eq. (\ref{Hamil}) is shown in Fig.~\ref{fig:fig1}. Individual data points are labeled according to which type of phase they belong to. For small $J_{2}$ and $J_{3}$, FM order prevails, however, with a diminished extent compared to its classical domain. For large $J_{2}$, and small to intermediate  $J_{3}$, the FM order gives way to stripe AF order, with quantum fluctuations extending its domain beyond the classically allowed region~\cite{Nagaev-1984}. Quantum effects also drastically shrink the domain of $(q,0)$ HM order to a small pocket. Upon increasing $J_{3}$ (for all $J_{2}$), the $(q,q)$ HM order onsets and prevails over the phase diagram. The real space illustration and the corresponding representative magnetic susceptibility profiles of the ordered phases are shown in Figs.~\ref{fig:fig1}(b)-(e), wherein the Bragg peaks of the respective types of magnetic orders are clearly resolved. Throughout the domain of both HM orders, we do not observe a discontinuous jump of the spiral wave-vector $q\to\frac{\pi}{2}$, thus pointing to the absence of commensurate magnetic orders with ${\bf Q}=(\pm\frac{\pi}{2},0)$ and $(\pm\frac{\pi}{2},\pm\frac{\pi}{2})$ as reported in Ref.~\cite{Sindzingre-2009}. The access to a continuous set of wave-vectors within our implementation of PFFRG together with a very large correlation area accounted for in the calculations, enables us to obtain a high-accuracy estimate of the shift in the spiral wave vectors with respect to the classical phases. Throughout the HM ordered phases it is found that quantum fluctuations always {\it increase} the magnitude of the wave vectors leading to more antiferromagnetic types of order, see Fig.~\ref{shift}. In the $(q,q)$ HM phase, the shift $\delta {\bf Q}$ has a maximal value of $\delta {\bf Q}\approx 37\%$ and decreases monotonically with increasing $J_{2}/|J_{1}|$ and $J_{3}/|J_{1}|$. Similarly, shifts of $\approx 100\%$ are found in portions of the classical $(q,0)$ HM phase that is turned into stripy AF order by quantum fluctuations, thus leading to the appearance of a ridge-like feature of $\delta {\bf Q}$ in the vicinity of the quantum phase boundary as seen in Fig.~\ref{shift}.

%=================PM===================
\begin{figure}
\includegraphics[width=1.0\columnwidth]{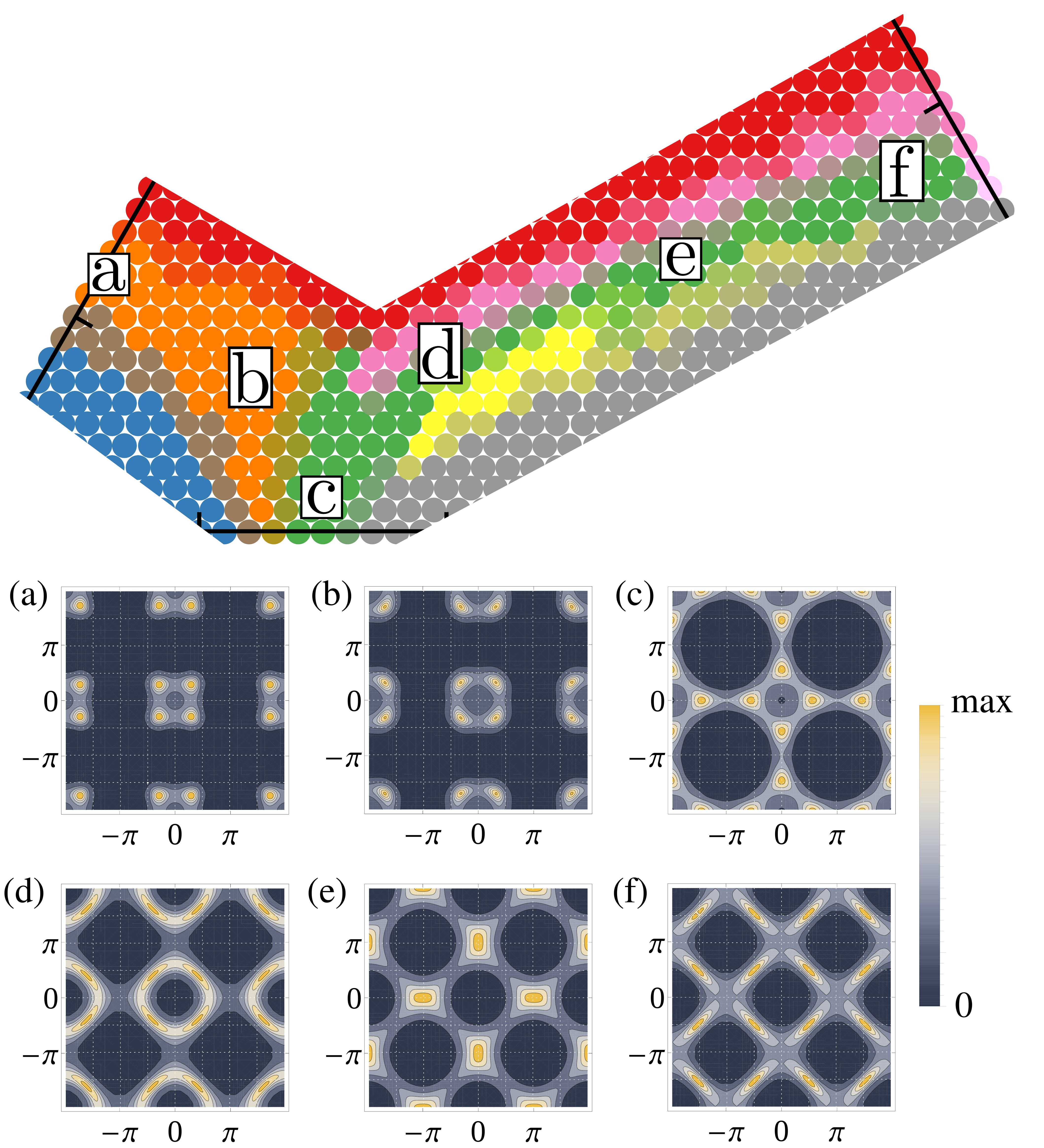}
\caption{\label{nemsus}
Magnetic susceptibility profiles in the PM regime of the quantum phase diagram evaluated at the corresponding labeled points.} 
\end{figure}
%====================================

The most salient effect of quantum fluctuations is the stabilization of an extended PM phase. Quantum fluctuations are found to eat up significant portions of the classical domains of the two HM phases, and to a comparatively lesser degree that of the FM phase (mostly at small $J_{3}$). This leads, in total, to a PM phase settling in the vicinity of most classical phase boundaries [see Fig.~\ref{fig:fig1}(a)]. In particular, on the $J_3=0$ line, a finite extent of the PM phase for $0.31(2) \leqslant J_{2}/|J_{1}|\leqslant 0.45(2)$ is found. This limit of the phase diagram has been previously addressed by a variety of methods with contrasting results on the issue of the presence of a paramagnetic phase, whose presence was first suspected in Ref.~\cite{Shannon-2004}. Exact diagonalization (ED) studies for up to $36$ spins, based on an analysis of the low-energy ED spectra argued for a PM phase for $0.4\lesssim J_{2}/|J_{1}|\lesssim 0.6$~\cite{Shannon-2006,Sindzingre-2007,Sindzingre-2009,Sindzingre-2010}. However, subsequent ED calculations for up to $40$ spins~\cite{Richter-2010}, based on the analysis of the ground state spin-spin correlation functions and the magnetic order parameter, found the stripe AF order to persist down till $J_{2}/|J_{1}| = 0.44$, but were inconclusive between the melting transition of the FM phase at $J_{2}/|J_{1}|=0.393$ and $J_{2}/|J_{1}|=0.44$. A high-order coupled cluster method (CCM) study claimed for the onset of stripe AF order immediately after the region of stability of the FM phase, thus finding no evidence for a PM phase~\cite{Richter-2010}. Finally, a variational Monte Carlo (VMC) study employing projected BCS wave-functions with spin-triplet pairing of spinons again found a nonmagnetic intermediate phase for $0.42\lesssim J_{2}/|J_{1}|\lesssim0.57$~\cite{Shindou-2011}. 

\begin{figure}
\includegraphics[width=1.0\columnwidth]{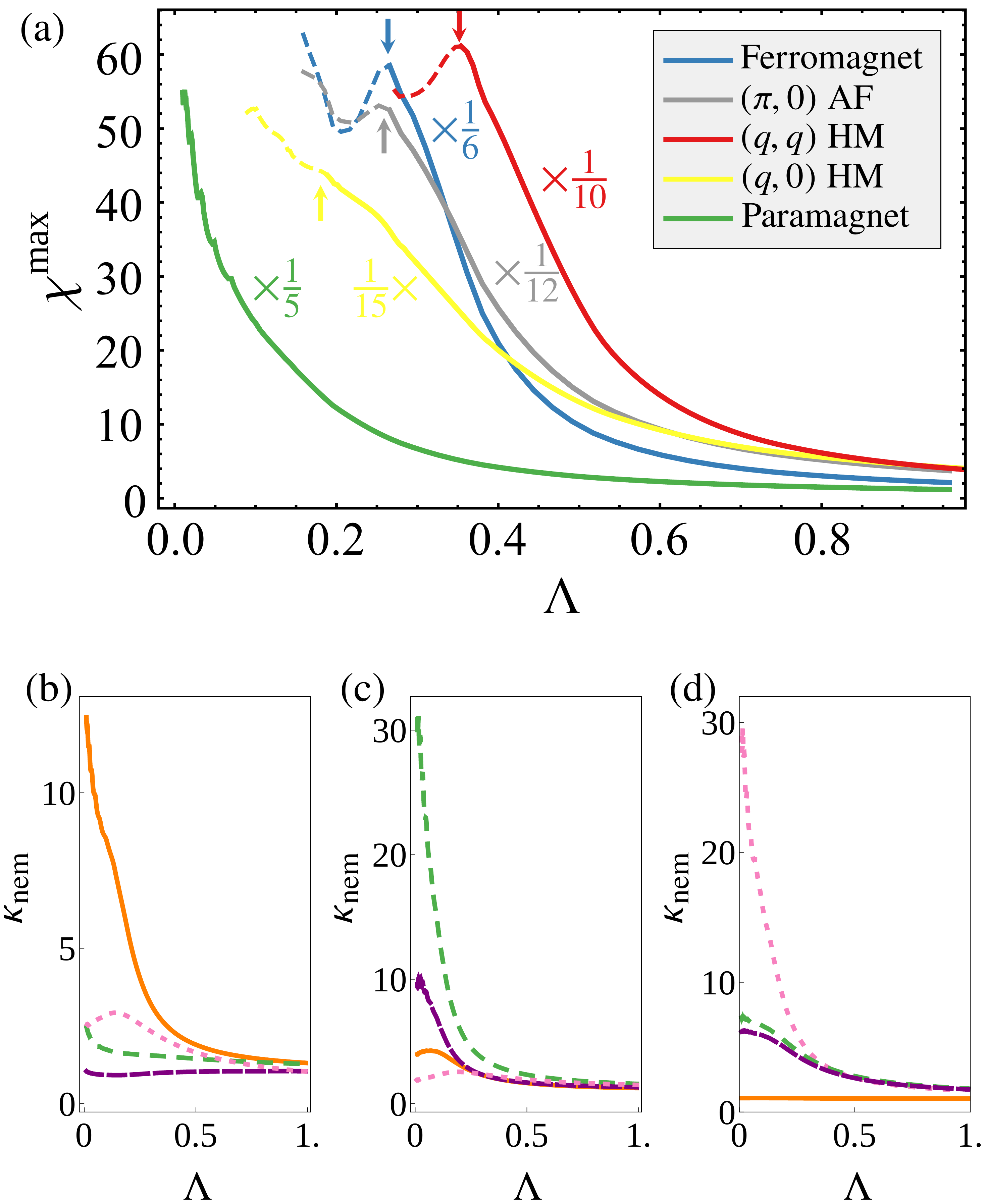}
\caption{\label{flow}
(a) Representative RG flows of the magnetic susceptibilities at the ordering wave vectors of the four ordered regimes of Fig.~\ref{fig:fig1} and the PM regime, evaluated at the following data points $(|J_{1}|,J_{2},J_{3})$: (i) FM: $(0.90,0.00,0.10)$, (ii) $(\pi,0)$: $(0.52,0.42,0.06)$, (iii) $(q,0)$ HM: $(0.46,0.42,0.12)$, (iv) $(q,q)$ HM: $(0.52,0.06,0.42)$, and (v) PM: $(0.66,0.26,0.08)$. The points at which the solid lines become dashed (marked by arrows) indicate an instability in the flow and express the onset of order. In the smooth flow (green curve) indicating paramagnetism, no such instability is found. (b)-(d) Representative nematic responses [Eq.~(\ref{nematic_susceptibility})] inside the three PM phases of Fig.~\ref{fig:fig1}(a), evaluated at the data points (0.74,0.14,0.12) (b), (0.68,0.30,0.02) (c), and (0.12,0.52,0.36) (d). The dark purple color curve (${\rm LN}_{\rm NN}$) corresponds to a lattice nematic bias on the NN bonds.} 
\end{figure}

In Fig.~\ref{nemsus}, we plot the magnetic susceptibility profiles at different parameter values in the PM region. Compared to the magnetic phases one observes a smearing of the spectral weight of the susceptibility with soft maxima at the Bragg peak positions of the nearest orders. Typical RG flow behaviors of the susceptibility in the different magnetically ordered and PM phases are shown in Fig.~\ref{flow}(a). While the PM flow does not show features of instability at any $\Lambda$ scale, the magnetic flows exhibit a pronounced kink below which the evolution of the susceptibility becomes numerically unstable. Note that the $(q,0)$ HM is characterized by a weak but still clearly resolved instability feature which manifests as a slight downturn of the susceptibility during the flow. This hints at small magnetic order parameters in this regime.

%==================Flow================
%====================================

At each point in the PM phase, we calculated the nematic response function $\kappa_\text{nem}$ for $s$-SN, $d$-SN, and LN orders. Interestingly, in a large portion of the PM phase, one response always clearly dominates over the other two, leading to a sharp distinction between nematic phases, see Fig.~\ref{fig:fig1}(a). Narrow intermediate regimes where responses are of similar size are indicated by interpolating colors in Fig.~\ref{fig:fig1}(a). A comparison of the responses shows that in the region surrounding the FM phase, i.e., when $J_{1}$ is dominant, the $s$-SN [see Fig.~\ref{bias}(b)] response undergoes the largest relative enhancement [see Fig.~\ref{flow}(b)], pointing to the existence of $s$-SN order in this regime [orange region in Fig.~\ref{fig:fig1}(a)]. As $J_{2}$ is increased, the $d$-SN [see Fig.~\ref{bias}(a)] response becomes dominant [see Fig.~\ref{flow}(c)]. This region is found to span a vast domain [green region in Fig.~\ref{fig:fig1}(a)] extending into the classical domain of the $(q,0)$ HM phase. In particular, the $d$-SN phase ranges down to the $J_3=0$ line as has previously been predicted by ED and VMC studies~\cite{Shannon-2006,Sindzingre-2007,Sindzingre-2009,Sindzingre-2010,Shindou-2009,Shindou-2011}. In a narrow strip between the $d$-SN and the $(q,q)$ HM phases, we observe strong LN responses [see Figs.~\ref{bias}(c) and \ref{flow}(d)] forming the pink region in Fig.~\ref{fig:fig1}(a). As mentioned before, the LN biasing pattern that was used to identify this phase acts on second neighbor couplings $J_2$. While in general, the breaking of the lattice-point group symmetry $C_{4v}$ down to $C_{2v}$ could also be probed with a nearest neighbor term of the form $\hat{\mathcal{H}}'_{\text{LN}}=\delta\sum_i(\hat{\bf S}_i\cdot\hat{\bf S}_{i+\hat{x}}-\hat{\bf S}_i\cdot\hat{\bf S}_{i+\hat{y}})$, the corresponding response is mostly found to be smaller [dark purple lines in Figs.~\ref{flow}(b)-\ref{flow}(d)]. As a result, the type of symmetry breaking that leads to the formation of the LN phase predominantly affects the correlations on diagonal bonds. This is expected because LN order parameters probe the {\it singlet} channel of two spins, which is energetically favored on antiferromagnetic bonds.

An interesting limit of the phase diagram is the line $J_1=0$ [right edge of the triangle in Fig.~\ref{fig:fig1}(a)] where only antiferromagnetic $J_2$ and $J_3$ interactions are finite. Here, the model reduces to two decoupled copies of the well-known Heisenberg model on the square lattice with antiferromagnetic NN and 2-NN couplings (which here correspond to $J_{2}$ and $J_{3}$ interactions, respectively). The existence of a paramagnetic phase between $J_3/J_2\approx0.4$ and $J_3/J_2\approx0.6$ is well established for this model~\cite{Read-1989,Gelfand-1989,Schulz-1992,Schulz-1996,Mambrini-2006,Richter-2010} and has also been confirmed by PFFRG [see Ref.~\cite{Reuther2010} and Fig.~\ref{fig:fig1}(a)]. The precise nature of this phase and the question whether it exhibits spontaneous symmetry breaking of valence-bond crystal (VBC)-type is, however, still debated~\cite{Jiang2012,Becca-2013,Gong2014,Danu-2016,Kumar-2011}. The most promising candidates for VBCs are columnar dimer order (with singlet dimers on the $J_2$ bonds, arranged in a columnar pattern) and plaquette order (with resonating dimers on square plaquettes of $J_2$ bonds). Previous PFFRG studies found that at $J_1=0$ both states yield only moderate and competing dimer responses such that the VBC scenario seems unlikely~\cite{Reuther2010}. To better understand the phase diagram at small $J_1$, we have performed additional PFFRG calculations also probing columnar and plaquette orders on the $J_2$ bonds. We generally find the VBC responses to be weakest throughout the PM phase. Even in the exact $J_1=0$ limit and for $J_{3}/J_{2}\gtrsim 0.55$, we find the LN responses to be about twice in magnitude compared to the columnar/plaquette VBC responses and an enhancement with increasing $J_{3}/J_{2}$, indicating that the LN might even survive in this limit. A similar observation is made for the $d$-SN phase which almost spreads out to the $J_1=0$ line (although $d$-SN order is not present at exactly $J_1=0$ due to vanishing ferromagnetic exchange). This indicates that a small $J_1$ perturbation away from the $J_1=0$ line might be sufficient to stabilize $d$-SN order.

\section{Discussion and conclusion}
In this work, we have investigated the spin-$\frac{1}{2}$ $J_{1}$-$J_{2}$-$J_{3}$ Heisenberg model on the square lattice with FM $J_1$ and AF $J_2$ and $J_3$ interactions. Using the PFFRG approach, we find that quantum fluctuations lead to the emergence of intertwined nematic orders over a vast region in parameter space. The analysis of generalized response functions yields different nematic domains hosting either $d$-SN, $s$-SN, or LN orders. We conclude that the $J_{1}$-$J_{2}$-$J_{3}$ Heisenberg model realizes the remarkable situation where three different types of nematic phases spontaneously arise within relative parametric proximity to each other.

Qualitatively, the locations of the nematic phases can be understood as follows. In regions where $J_1$ is the dominant coupling but $J_2$ and $J_3$ are strong enough to melt the FM order, the remaining FM correlations generate resonating nearest neighbor triplets leading to spin nematic order. The isotropy of FM order in real space persists in the nematic state giving rise to an $s$-SN phase at small $J_2$ and $J_3$. In the same way, the FM $J_1$ coupling induces spin nematic order in parts of the paramagnetic regime located near the $(\pi,0)$ AF or $(q,0)$ HM phases. However, the $(\pi,0)$ AF and $(q,0)$ HM orders break the fourfold rotation symmetry of the square lattice. This symmetry breaking persists in the corresponding melted state leading to $d$-SN order in the vicinity of the $(\pi,0)$ AF and $(q,0)$ HM phases, which agrees with the findings in Ref.~\cite{Sindzingre-2010}. Finally, the narrow strip of LN order close to the $(q,q)$ HM phase can be described as a melted version of the classical $(q,q)$ HM state. The close relation between LN states and coplanar spiral phases has already been realized in other frustrated spin models~\cite{Hwang2013,Mulder2010}, and is attributed to the fact that both states break the fourfold lattice-rotation symmetry. It is important to note that the $(q,q)$ HM state is the only type of classical ground state in this model where the ordering wave vector does not have at least one vanishing component, indicating that FM correlations due to $J_1<0$ are less important for the formation of this phase. Consequently, the LN phase in the vicinity of the $(q,q)$ HM ordered region does not exhibit nematicity in spin space. The absence of spin nematicity in this part of the phase diagram has previously been realized in Ref.~\cite{Sindzingre-2010}.

Methodologically, we have demonstrated that the PFFRG is a suitable tool to study spin and lattice nematic states in frustrated quantum magnets and to resolve the complex interplay between these phases. An extended version of our approach could be used to study antisymmetric spin-nematic states of $p$-type. We leave such investigations for future work.

\begin{acknowledgments}
We thank M. S. Laad and I. I. Mazin for useful discussions. The work was supported by the European Research Council through ERC-StG-TOPOLECTRICS-Thomale-336012. Y.I. and R.T. thank the DFG (Deutsche Forschungsgemeinschaft) for financial support through SFB 1170. R.N. acknowledges funding from the Visitor program at APCTP and also acknowledges support through NRF funded by MSIP of Korea (2015R1C1A1A01052411). P.G. acknowledges CSIR (India) for financial support. B.K. acknowledges the financial support under UPE-II scheme of JNU, DST-PUSRE and DST-FIST support for the HPC facility in SPS, JNU. J.R. is supported by the Freie Universit\"at Berlin within the Excellence Initiative of the German Research Foundation. Y.I. gratefully acknowledges the Gauss Centre for Supercomputing e.V. for funding this project by providing computing time on the GCS Supercomputer SuperMUC at Leibniz Supercomputing Centre (LRZ). P.G. and B.K. acknowledge IUAC (India) for using the HPC facility. 
\end{acknowledgments}

Y.I. and P.G. are equally contributing first authors.

%\bibliography{square.bib}


\end{document}